\documentstyle[aps]{revtex}
\begin{document}
\draft
\title{Scanning tunneling microscopy and spectroscopy at low temperatures of the (110) surface of Te doped GaAs single crystals}
\author{A. Depuydt and C. Van Haesendonck}
\address{Laboratorium voor Vaste-Stoffysica en Magnetisme, Katholieke Universiteit Leuven, B-3001 Leuven, Belgium}
\author{N.S. Maslova, V.I. Panov, and S.V. Savinov}
\address{Chair of Quantum Radio Physics, Moscow State University, 119899 Moscow, Russia}
\author{P.I. Arseev}
\address{Physical Institute, Moscow State University, 117924 Moscow, Russia}
\date{\today}
\maketitle
\begin{abstract}
We have performed voltage dependent imaging and spatially resolved spectroscopy on the (110) surface of Te doped GaAs single crystals with a low temperature scanning tunneling microscope (STM). A large fraction of the observed defects are identified as Te dopant atoms which can be observed down to the fifth subsurface layer. For negative sample voltages, the dopant atoms are surrounded by Friedel charge density oscillations. Spatially resolved spectroscopy above the dopant atoms and above defect free areas of the GaAs (110) surface reveals the presence of conductance peaks inside the semiconductor band gap. The appearance of the peaks can be linked to charges residing on states which are localized within the tunnel junction area. We show that these localized states can be present on the doped GaAs surface as well as at the STM tip apex.
\end{abstract}
\section{Introduction}
\begin{par}
Since its introduction, scanning tunneling spectroscopy (STS) was expected to reveal the electronic structure of surfaces with a spatial resolution of the order of the interatomic distance~\cite{Binnig}. When combined with scanning tunneling microscopy (STM) imaging, STS is a powerful tool for studying the influence of lattice defects and impurities on the local electronic structure. Magnetic impurities in normal metals~\cite{Depuydt1} as well as superconductors~\cite{Yazdani} or defects in semiconducting materials~\cite{Domke,Feenstra1,Maslova1,Depuydt2} have been studied. 
\end{par}
\begin{par}
The surface of the III-V compound semiconductor GaAs has been investigated extensively. STM and STS measurements have provided detailed information on the local electronic structure of dopant atoms and other atomic scale defects. Johnson {\em et al.}~\cite{Johnson} showed by voltage dependent STM imaging at room temperature that substitutional Zn and Be dopant atoms (acceptors) at the surface and in the upper subsurface layers influence the electronic structure of the GaAs (110) surface. Zheng {\em et al.}~\cite{Zheng} and Domke {\em et al.}~\cite{Domke} obtained similar results for substitutional Si dopant atoms (donors). At low temperatures, Van der Wielen {\em et al.}~\cite{van der Wielen} observed Friedel charge density oscillations around Si dopant atoms near the GaAs surface. Other groups have reported on the presence of atomic scale defects at the cleaved GaAs (110) surface by STM at room temperature. The geometric and electronic structure of As vacancies~\cite{Domke,Lengel,Chao}, dopant-vacancy complexes~\cite{Domke} and As antisite defects~\cite{Feenstra1} have been identified.
\end{par}
\begin{par}
The experienced difficulties of reproducibility indicate, however, that the application of combined STS and STM measurements requires a detailed knowledge of the relevant physical processes governing the behavior of nanoscale tunnel junctions. For nanoscale junctions the local density of states in the contact area is strongly altered by tip-sample interactions. These interactions result in considerable shifts of the allowed energy levels, where deep lying levels may be driven through the Fermi level~\cite{Maslova2,Mallet,Wildoer}. Localized states can not only be connected with the sample surface~\cite{Maslova1}, but also appear at the tip apex~\cite{Vazquez}. Because the radii of the localized states are of the order of the contact area, the tunneling current in STM and STS experiments can be dominated by electron transport through one single localized state~\cite{Renner}. Therefore, we have to take into account the finite relaxation rate of the electrons which occupy the localized states. At low temperatures the relaxation rate may become comparable to the tunneling rate for the electrons which will be driven out of equilibrium~\cite{Agam}. The low relaxation rate will give rise to the appearance of localized charges in the contact area which strongly influence the position of the localized states with respect to the Fermi energy~\cite{Depuydt2}.
\end{par}
\begin{par}
In the present paper, we present the results of our low temperature STM/STS investigation of Te doped GaAs crystals which are cleaved along the (110) plane. By STM imaging at different voltages and polarities, we are able to distinguish different types of defects which can be located in different subsurface layers. The main type of defect is identified as an n-type substitutional dopant atom, ${\rm Te}_{\rm As}$, i.e., a Te atom occupying an As lattice site. The STM images of the dopant atoms depend on the applied sample voltage. We have also obtained spatially resolved spectroscopy curves for different positions of the tip in the vicinity of a Te impurity. The differential conductance curves show the presence of peaks inside the semiconductor band gap. The position and height of the peaks depend on the position of the tip with respect to the impurity. We will argue that both the topographic STM images and the conductivity curves reflect the presence of charges which are localized in the tunnel junction area. As already indicated above, the localized charges are the combined result of the presence of localized states within the nanoscale tunnel junction area and the non-equilibrium electron distribution which is caused by the low relaxation rate at low temperatures. Differential conductance curves taken above a defect free surface area with different STM tips confirm that also the tip contains localized states which can be charged. The appearance and position of the conductance peaks within the band gap can be linked to the complicated voltage dependence of the charging effects. Finally, at negative sample voltages we clearly observe Friedel charge density oscillations around the ionized Te dopant atoms. 
\end{par}
\section{Experiment}
\begin{par}
The STM and STS data have been obtained with a home built low temperature STM with an in situ cleavage mechanism~\cite{Oreshkin}. The samples are n-type GaAs single crystals which are doped with Te. The nominal concentration of Te atoms is $5 \times 10^{17} \, {\rm cm}^{-3}$. It is well known that Te acts as a donor dopant atom which occupies an As lattice position~\cite{Madelung}. At our relatively low doping level, compensation effects~\cite{Domke} may be neglected. The crystals are cleaved along the (110) plane after cooling down to liquid helium temperature. The partial vapor pressure of oxygen is extremely low at this temperature, implying that surfaces like the GaAs (110) surface will stay atomically clean for many days. All the STM and STS measurements are done with $\rm Pt_{\rm 80}$$\rm Ir_{\rm 20}$ tips, cut ex situ with scissors.
\end{par}
\begin{par}
Our samples contain ohmic contacts obtained by thermodiffusion which allow to perform electrical transport measurements. From Hall measurements we have determined that the density of electrical carriers at $5 \, {\rm K}$ is $8.9 \times 10^{17} \, {\rm cm}^{-3}$, which is sufficient to result in metallic conductivity at low temperatures~\cite{Emelyanenko}. Te is a shallow impurity for GaAs, i.e., the Te atom occupying an As lattice site provides a 5s electron which is weakly bound to a positively charged Te ion. The localization radius for such a 5s valence electron is about $7 \, {\rm nm}$. At doping levels exceeding $4.5 \times 10^{17} \, {\rm cm}^{-3}$ the orbitals from neighboring doping atoms will overlap, providing metallic conductivity with a Fermi level close to the edge of the conduction band. The metallic behavior of our samples is confirmed by the temperature dependence of the conductivity $\sigma(T) - \sigma(T \rightarrow 0) \propto \sqrt{T}$~\cite{LR} with an extrapolated conductivity $\sigma(T \rightarrow 0) \simeq 1200 \, {\Omega}^{-1} {\rm cm}^{-1}$. The high quality of our GaAs crystals also allows the observation of Shubnikov-de Haas oscillations in the magnetic field dependence of the low temperature conductivity. 
\end{par}
\section{Experimental results}
\begin{par}
The (110) GaAs surface has a simple (1x1) structural relaxation which leads to surface states located outside of the semiconductor band gap. Unoccupied Ga and occupied As surface states can be found above and below the band gap, respectively. Therefore, the STM image of the clean (110) GaAs surface taken at positive and negative sample voltages is within a first order approximation determined by the Ga and As sublattices, respectively~\cite{Feenstra2}. This is illustrated in the inset of Fig.~\protect\ref{Fig.1} which shows two STM pictures of the GaAs(110) surface at opposite polarity of the tunneling voltage. A high doping level and/or voltage dependent band bending~\cite{Feenstra3} can result in the presence of occupied states in the conduction band. In that case, depending on the applied voltage, both the Ga and the As sublattice will contribute to the STM images~\cite{Jiang}.
\end{par}
\begin{par}
Figure~\protect\ref{Fig.1} shows the current versus voltage characteristic above an atomically flat area of the GaAs (110) surface. Due to the voltage dependent band bending, the measured band gap tends to be larger than the bulk value~\cite{Yoshita} (For GaAs ${E_g \simeq 1.52 \ \rm eV \ at \ 5 \rm K}$, ${E_g \simeq 1.43 \ \rm eV \ at \ 300 \rm K}$~\cite{Madelung}) and at low temperatures this difference can become quite large~\cite{Maslova1}. Voltage dependent band bending is the result of the space charge region at the surface of the sample which compensates the electric field between the tip and the sample~\cite{Feenstra3}. We will show that at low temperatures this band bending is very sensitive to localized charges which are induced in the STM contact area.
\end{par}
\begin{par}
Figure~\protect\ref{Fig.2} shows an STM topographic image of the cleaved GaAs (110) surface at a sample voltage of $-1.5 \, {\rm V}$. We clearly distinguish different types of defects superimposed on the atomic lattice~\cite{van der Wielen}. We will restrict ourselves to the investigation of one type of defect, referred to as the A-type defect. Figures~\protect\ref{Fig.3}(a) and 3(b) show one A-type defect at different values of the sample voltage. This defect is observed as a round hillock feature which at negative polarity of the sample voltage is surrounded by a darker ring. As will be discussed in more detail in Section V, the ring like features can be interpreted in terms of Friedel charge density oscillations which result from the screening of charged defects. The statistical distribution of the charged A-type defects within the different subsurface layers (see below) indicates that these defects correspond to the Te dopant atoms. On the other hand, we note that the two STM images shown in Figs.~\protect\ref{Fig.3}(a) and 3(b) look very similar to voltage dependent STM images of substitutional Si$_{\rm Ga}$ dopant atoms for the GaAs (110) surface~\cite{Domke,Zheng,van der Wielen}. We conclude that the A-type defects appearing in Fig.~\protect\ref{Fig.2} are substitutional Te$_{\rm As}$ dopant atoms which occupy As lattice sites. 
\end{par}
\begin{par}
Figure~\protect\ref{Fig.3}(c) is an STM image at a sample voltage of $+0.5 \, {\rm V}$ of the same ${\rm Te}_{\rm As}$ dopant atom shown in Figs.~\protect\ref{Fig.3}(a) and 3(b). While scanning the area surrounding the dopant atom, the image of the dopant atom suddenly switched from a hillock feature to a depression (the scanning direction is downwards). The different topography can not simply be the result of a double tip effect or any other mechanical instability, since we clearly observe continuous atomic rows on the flat surface at the left and right hand side of the dopant atom. When increasing the sample voltage to $+1.5 \, {\rm V}$, the dopant image continues to appear as a depression (see Fig.~\protect\ref{Fig.3}(d)). We note that the switching of the contrast in Fig.~\protect\ref{Fig.3}(c) only affects the imaging at positive sample voltages, i.e., the Friedel charge density oscillations at negative sample voltages continue to appear as shown in Fig.~\protect\ref{Fig.3}(a). The theoretical model, which will be introduced in Section IV, allows to link the switching of the contrast in Fig.~\protect\ref{Fig.3}(c) to a change of the localized charges residing on the tip apex and/or the dopant atom.
\end{par}
\begin{par}
The different intensities for the A-type defects in Fig.~\protect\ref{Fig.2} are caused by the fact that the Te$_{\rm As}$ dopant atoms can be observed in different layers below the surface~\cite{Johnson,Zheng}. According to Zheng {\em et al.}, dopant atoms at the surface are expected to behave differently from the dopant atoms in the subsurface layers~\cite{Zheng}. According to the statistical distribution, the A-type dopant atoms can be identified as subsurface dopant atoms, while the B-type defects in Fig.~\protect\ref{Fig.2} probably correspond to dopant atoms residing at the GaAs surface. The inset of Fig.~\protect\ref{Fig.4} shows the height profiles across three A-type dopant atoms located in different layers. The observed corrugations can be grouped in five categories corresponding to the top five subsurface layers. In Fig.~\protect\ref{Fig.4} we have plotted on a logarithmic scale the average corrugation for the five layers. The average corrugation depends exponentially on the depth below the surface. The number of visible layers corresponds to what other authors have reported for measurements at room temperature~\cite{Johnson,Zheng}. On the other hand, we note that Zheng {\em et al.}~\cite{Zheng} have observed at room temperature a linear dependence of the corrugation height on the layer number for Si dopant atoms. The exponential dependence at low temperatures may be caused by a spatial localization of the probed electron states which becomes more pronounced with decreasing temperature. 
\end{par}
\begin{par}
Figure ~\protect\ref{Fig.5}(a) shows the normalized conductance curves $(dI/dV)/(I/V)$ which have been obtained in the vicinity of a Te$_{\rm As}$ dopant atom. The curves have been averaged within the indicated three square areas around a Te$_{\rm As}$ dopant atom located in the first subsurface layer. After averaging about 70 measured $I(V)$ curves within one area, the differential conductance $dI/dV$ is obtained numerically and normalized. Several larger and smaller peaks can be observed inside the semiconductor band gap. The position as well as the intensity of these peaks strongly depends on the surface area which is being probed. We will show that the peaks reflect the presence of localized states which can be associated with defects and/or the tip apex. When compared to room temperature, the influence of localized states becomes more dominant in the tunneling process at low temperatures because of the low relaxation rate for the electrons. The relevance of states localized at the tip apex is illustrated by the two normalized conductance curves in Fig.~\protect\ref{Fig.5}(b). The data have been obtained above an atomically flat, defect free area on the GaAs surface with two different STM tips. The presence of a peak in the tunneling conductance near the band gap edge is obvious for one tip, while this peak is absent for the other tip used in the experiment. The fact that two different STM tips can result in different conductance curves is consistent with a change of the charge localized at the tip apex. A theoretical model which takes into account such charging effects is presented in Section IV below. 
\end{par}
\begin{par}
As discussed in Section II, the doping with Te atoms provides a metallic conductivity in our GaAs samples even at low temperatures. The ionized Te atoms correspond to a localized charge which is screened by the conduction electrons. This gives rise to the presence of Friedel charge density oscillations which became already visible in Fig.~\protect\ref{Fig.2} and in Fig.~\protect\ref{Fig.3}. Figure~\protect\ref{Fig.6} shows a detailed STM image of a ${\rm Te}_{\rm As}$ dopant atom on the GaAs (110) surface taken at a sample voltage of $-1.5 \, {\rm V}$. In order to highlight the presence of the Friedel charge density oscillations, the atomic corrugation of the image has been filtered out digitally. The ring like structures are very similar to the Friedel oscillations appearing on the GaAs (110) surface around Si dopant atoms~\cite{van der Wielen}. As discussed in more detail in Section V, the image shown in Fig.~\protect\ref{Fig.6} can not be simply related to the standard screening model for the bulk material~\cite{Hofmann}. According to this model, the oscillation period should be given by half the bulk Fermi wave vector: $\lambda_{\rm F}/2 \simeq 10.5 \, {\rm nm}$ for the carrier density obtained from the Hall effect (see Section II), while the oscillation period inferred from Fig.~\protect\ref{Fig.6} is $3.3 \, {\rm nm}$.
\end{par}
\section{Theoretical model}
\begin{par}
The conductivity of a tunnel junction between a semiconductor and a metal tip is usually calculated by relying on the concept of band bending~\cite{Feenstra3}. The position of the energy bands is determined by the electrostatic potential which is related to the charge density through Poisson's equation. Integration of Poisson's equation gives the band bending in the semiconductor. An essential assumption is that the Fermi-Dirac statistics can be used to determine the equilibrium electron occupation numbers from which the charge density distribution is obtained. In a nanoscale tunnel junction, the current can be influenced and even be dominated by tunneling through localized states which can be present at the semiconductor surface as well as at the tip apex~\cite{Vazquez}. At low temperatures we have to take into account the finite relaxation rate of the electrons occupying the localized states. The electron distribution will be out of equilibrium and an additional charge can appear in the tunnel contact area due to the presence of the localized states~\cite{Maslova1}. This additional charge will influence the local electronic spectrum. In order to describe this influence we propose a theoretical model which is a generalisation of previously used approaches and includes both non-equilibrium effects and the influence of localized states in the tunnel contact area.
\end{par}
\begin{par}
The influence of an additional localized charge on the tunneling characteristics has to be treated self-consistently. It is necessary to take into account both a Hubbard type repulsion between localized electrons as well as the electrostatic potential at the semiconductor surface due to the localized charge. This is imposed by the fact that both the typical tip-sample separation and the typical radius of a localized state are comparable to the inter-atomic distance. The Coulomb interaction of the Hubbard type shifts the energy of the localized state by an amount ${U \simeq e^2/a_{0}}$, where $a_{0}$ is the radius of the localized state. Next, a potential $W$ has to be introduced to describe the interaction of the electrons at the semiconductor surface with the additional charge present on the localized state~\cite{Arseev1}. In general, the exact calculation of the Coulomb potential at a semiconductor surface is very complicated because one needs to know both the geometry of the tunnel contact and the distribution of the electric field in the contact area. There are two limiting cases for which the calculations can be simplified, but are still able to reproduce the main characteristic features of the tunneling conductivity and the STM imaging. In the limit of strong screening, when the effective radius of the potential is of the order of the inter-atomic distance, one can treat $W$ as a point like potential. In the limit of weak screening, when the effective radius of the potential is much larger than the inter-atomic distance and the tunnel contact size, the potential $W$ at the semiconductor surface stays approximately constant in the vicinity of the contact. Here, we will restrict ourselves to the case of strong screening.
\end{par}
\begin{par}
The extra charge residing on the localized states and the tunneling conductivity can be obtained from a self-consistent approach based on the diagram technique for non-equilibrium processes~\cite{Arseev2}. In order to describe the tunneling processes for an STM junction in the presence of a localized state, we use a model which includes three subsystems: an ideal semiconductor, a localized electronic level (connected with a surface defect or with a tip apex state) and a normal metal (the STM tip). The subsystems are connected by tunneling matrix elements. We add an interaction of the semiconductor electrode with a thermal bath in order to take into account a finite relaxation rate for the electrons. We consider such a finite relaxation rate only for the semiconductor electrons: The electrons in the metallic electrode (STM tip) are assumed to be in thermal equilibrium.
\end{par}
\begin{par}
Figure~\protect\ref{Fig.7} gives a schematic view of the tunnel junction we are considering. The thermal bath is connected to the semiconductor via a relaxation rate $\gamma$. The expression for the tunneling conductivity turns out to be less sensitive to the details of the connection (point-like connection or a distribution of scattering centers)~\cite{Arseev2}. The initial position of the localized state in the absence of any tip-sample interaction corresponds to an energy $\varepsilon_{d}^{0}$. The STM contact induces a shift of the localized state towards a voltage dependent energy $\varepsilon_{d}$. We want to stress that in our model the localized state can in principle be any state which is localized within the tunnel junction area. In our STM/STS measurements the localized state can, e.g., be a surface impurity state or a state localized at the tip apex. In the example illustrated in Fig.~\protect\ref{Fig.7} the localized state has been assigned to the tip apex and is connected to the bulk of the metallic tip metal via a relaxation rate $\gamma_{0}$. The localized state is connected to the semiconductor electrode via the tunneling rate $\Gamma$. 
\end{par}
\begin{par}
Figure~\protect\ref{Fig.8} shows some typical results of our numerical evaluation of the analytical expression for the tunneling conductivity. Each curve corresponds to a set of typical values for the relaxation rates $\gamma$ and $\gamma_{0}$, the tunneling rate $\Gamma$ and the initial position $\varepsilon_{d}^{0}$ of the localized state. All the relevant parameters are expressed in units of the energy $\Delta$ which corresponds to half the semiconductor band gap, i.e., $E_{g}/2$ (see Fig.~\protect\ref{Fig.7}). The initial model density of states of the semiconductor is shown by the dotted line in Fig.~\protect\ref{Fig.8}. As indicated in Fig.~\protect\ref{Fig.7}, the bands are assumed to have a width $4 \Delta$. The on-site Hubbard repulsion ${U \sim e^2/a_0}$ is about $0.5 - 1 \, {\rm eV}$ for a localization radius $a_{0} \simeq 0.5-1 \, {\rm nm}$. For our calculations we have taken $U = \Delta$. A similar choice $W = \Delta$ has been made for the semiconductor surface Coulomb potential. The qualitative features of the tunneling conductivity are insensitive to variations of the Coulomb parameters $U$ and $W$ for reasonable choices of these parameters. For the numerical evaluation two different situations have been investigated: (i) The initial position $\varepsilon_{d}^{0}$ of the localized state is inside the band gap (Fig.~\protect\ref{Fig.8}, curve 1), and (ii) the initial position $\varepsilon_{d}^{0}$ of the localized state is in the conduction or the valence band (Fig.~\protect\ref{Fig.8}, curves 2 and 3). 
\end{par}
\begin{par}
The calculated conductivity curves obviously differ from the standard tunneling conductivity curves which are expected for STS measurements. Our calculations clearly reveal a shift of the band gap edges which becomes more pronounced when decreasing the relaxation rates (Fig.~\protect\ref{Fig.8}, curves 2 and 3). The non-equilibrium electron distribution leads to a charge accumulation on the localized state with initial energy $\varepsilon_{d}^{0}$. Due to the Coulomb repulsion this results in a shift of the level to a position $\varepsilon_{d}$, where the shift $\varepsilon_{d} - \varepsilon_{d}^{0}$ is comparable to the value of the band gap $E_{g} = 2 \Delta$. Despite its initial energy, the localized state can emerge as a peak near the band gap edge (Fig.~\protect\ref{Fig.8}, curve 1). Near the band gap edge, the tunneling current rapidly grows with increasing tunneling voltage, implying major changes in the charge residing on the localized state. The exact location of the conductance peak is sensitive to variations of the parameters $\Gamma$, $\gamma$, $\gamma_0$ and $\varepsilon_d^{0}$, which determine the value of the induced charge. According to our calculations, the peak is not influenced by the position of the Fermi level relative to the band gap edges.
\end{par}
\begin{par}
Our theoretical model can be generalized to the case where there exist several localized states which are connected with the STM tip apex and/or with a defect in the tunnel junction area. Taking into account the induced charges connected with all the localized states, one expects to observe several peaks in the tunneling conductivity. Finally, from our theoretical analysis we can draw the important conclusion that a peak in the tunneling conductance can also appear above an atomically flat surface area, provided a localized state is present at the apex of the STM tip.
\end{par}
\section{Discussion}
\begin{par}
Our low temperature STM/STS study of the GaAs (110) surface reproduces several of the basic features which have been reported for room temperature experiments. These features include the imaging of the Ga and As sublattice as well as the possibility to identify dopant atoms and other atomic scale defects. On the other hand, our low temperature experiments reveal specific features which can not be observed or are less pronounced at room temperature. As discussed in Section IV, an unusual behavior of the tunneling conductivity in low temperature STM/STS experiments can be associated with the presence of localized states in the nanoscale junction area. While our model can not provide a complete, quantitative understanding of the results, we are able to provide a qualitative understanding of the specific features which appear at low temperatures. 
\end{par}
\begin{par}
The charge residing on localized states associated with surface impurities causes a local change of band bending, which leads to the observed contrast when imaging a dopant impurity (see Fig.~\protect\ref{Fig.3}). Our theoretical model also indicates that a charge associated with a localized state at the tip apex can have a strong influence on the image contrast. Due to the Coulomb interaction, the energy of a localized state in the tunnel junction area will be shifted when the charge on the localized state changes. Therefore, the STM image contrast also depends on the modification of the initial electronic spectrum by the extra localized charge. This implies that a self-consistent treatment of the tunneling process (see Section IV) is required to understand the voltage dependence of the contrast when imaging an impurity (see Fig.~\protect\ref{Fig.3}). The details of this voltage dependence can obviously be different for different experiments. 
\end{par}
\begin{par}
The sudden change of the contrast in Fig.~\protect\ref{Fig.3}(c) can be explained in terms of a change in the charge which is localized at the tip apex or on the impurity atom. On the other hand, in the absence of any sudden variation in the localized charges, the voltage dependence of the contrast is completely reproducible.
\end{par}
\begin{par}
As illustrated in Fig.~\protect\ref{Fig.5}, the experimental results for the tunneling conductivity strongly depend on the investigated area as well as on the tip which is being used. In Section IV we have indicated that the presence of several peaks in the tunneling conductivity can be associated with the influence of several localized states. These localized states may be associated with a dopant atom or with the tip apex, but the GaAs surface states can also result in a set of additional states with initial energies lying in the conduction or the valence band. Charging effects are able to shift these surface states into the band gap and give rise to peaks in the tunneling conductivity. Because the charge accumulated on a localized state is determined by the relaxation and tunneling rates, it will also depend on the tip-sample and the tip-defect distance. By changing the STM tip position, one can obtain tunneling conductance curves where the position and the height of the peaks can be very different. The influence of the tip position on the conductance peaks is illustrated in Fig.~\protect\ref{Fig.5}(a). As shown in Fig.~\protect\ref{Fig.5}(b), a conductance peak can also be present for an atomically flat surface. According to our theoretical model such a peak can be directly related to the charge residing on a localized state at the tip apex. We note that the experimental conductance curves shown in Fig.~\protect\ref{Fig.5} also allow to verify the shift of the semiconductor band gap edges which is predicted by our model. 
\end{par}
\begin{par}
In our experiments Friedel charge density oscillations can only be observed at negative sample voltages. The absence of the oscillations at positive voltages is caused by band bending which results in a depletion of the surface area and a shift of the Fermi level away from the conduction band \cite{van der Wielen}. Electrons, which are able to screen the positive charge of the ionized Te dopant atoms, are only present at negative sample voltages. As mentioned in Section III, the Friedel oscillations shown in Fig.~\protect\ref{Fig.6} can not be simply explained in terms of the standard screening model. This is not surprising, since the image shown in Fig.~\protect\ref{Fig.6} is obtained at one particular sample voltage. The standard model, which predicts that the oscillation period is given by $\lambda_{F}/2$, takes into account conduction electrons with all possible energies \cite{Hofmann}. A detailed fitting of the observed Friedel oscillations requires an exact knowledge of the two-dimensional band structure of the GaAs surface. Additional complications arise because the tip-sample separation (about $0.5 \, {\rm nm}$) does not exceed the width of the ring like structures around the Te impurity (more than $2 \, {\rm nm}$ for GaAs). Therefore, charged states at the tip apex are likely to modify the distribution of the electron density and the corresponding STM image of the Friedel oscillations. On the other hand, the number of screening electrons below the Fermi level in the surface region depends on the local band bending which is determined both by the applied voltage and the presence of localized charges in the tunnel junction area (see Section IV). Finally, the distance between neighboring dopant atoms is comparable to the screening length. This will result in a superposition of the Friedel oscillations caused by different dopant atoms. The fact that the Friedel oscillations shown in Fig.~\protect\ref{Fig.6}(a) tend to be deformed at larger distances, is probably related to this superposition of oscillations. 
\end{par}
\section{Conclusion}
\begin{par}
We have studied n-type GaAs single crystals which are doped with Te atoms. The electrical transport properties reveal a metallic behavior of the GaAs crystals down to liquid helium temperatures. We have performed voltage dependent imaging and spatially resolved spectroscopy on the (110) surface of the in situ cleaved crystals by means of a low temperature scanning tunneling microscope. The larger fraction of the atomic scale defects are identified as substitutional ${\rm Te}_{\rm As}$ dopant atoms. These dopant atoms can be observed in the surface layer as well as in the next four subsurface layers and become surrounded by Friedel charge density oscillations at negative sample voltages.
\end{par}
\begin{par}
We have developed a theoretical model which qualitatively accounts for the voltage dependent contrast of the STM topographic images. The model also provides an explanation for the conductance peaks which appear in the semiconductor band gap and appear very differently when changing the tip position. Our model is based on the presence of charges residing on localized states in the tunnel junction area. The charges appear because of the non-equilibrium electron distribution in the STM contact area which results from a finite relaxation rate for the electrons at low temperatures. The localized charges are not only associated with the Te dopant atoms or with other atomic scale defects, but also appear on states which are localized at the tip apex. The presence of localized states at the tip apex allows to understand tip dependent anomalies which can even be observed on atomically flat surface areas.
\end{par}
\acknowledgments
The work at the K.U.Leuven has been supported by the Fund for Scientific Research - Flanders (FWO) as well as by the Flemish Concerted Action (GOA) and the Belgian Inter-University Attraction Poles (IUAP) research programs. The collaboration between Moscow and Leuven has been funded by the European Commission (INTAS, project 94-3562). The work in Moscow has been supported by the Russian Ministry of Research (Surface atomic Structures, grant 95-1.22; Nanostructures, grant 1-032) and the Russian Foundation of Basic Research (RFBR, grants 96-0219640a and 96-15-96420). We are much indebted to I.~Gordon for performing the electrical transport measurements.

\newpage
\begin{figure}
\protect\caption{Tunneling current as function of sample voltage measured above a defect free area of the GaAs (110) surface. $E_{g}$ indicates the estimated width of the semiconductor band gap. The inset shows two STM images of the GaAs (110) surface at different polarity of the sample voltage. The tunneling current is fixed at $20 \, {\rm pA}$ for the two images.}
\label{Fig.1}
\end{figure}
\begin{figure}
\protect\caption{Typical topographic image of the GaAs (110) surface which includes many atomic scale defects. Subsurface ${\rm Te}_{\rm As}$ dopant atoms have been labeled with the letter A. The defects labeled with the letter B probably correspond to surface ${\rm Te}_{\rm As}$ atoms. The scanned area measures $41 \, {\rm nm} \times 41 \, {\rm nm}$ and the grey scale corresponds to $0.5 \, {\rm nm}$ height variations between black and white. The sample voltage is $-1.5 \, {\rm V}$ and the tunnel current has been fixed at $60 \, {\rm pA}$.}
\label{Fig.2}
\end{figure}
\begin{figure}
\protect\caption{STM images of a ${\rm Te}_{\rm As}$ dopant atom at different values of the sample voltage: (a) $-1.5 \, {\rm V}$, (b) $+1 \, {\rm V}$, (c) $+0.5 \, {\rm V}$, and (d) $+1,5 \, {\rm V}$. The scanned area measures $5.8 \, {\rm nm} \times 5.8 \, {\rm nm}$. The tunnel current is fixed at $20 \, {\rm pA}$ and the scanning direction is downwards.}
\label{Fig.3}
\end{figure}
\begin{figure}
\protect\caption{Exponential dependence of the average corrugation height on the subsurface layer number. The inset shows the filtered cross sections through three ${\rm Te}_{\rm As}$ dopant atoms which are observed in the upper three subsurface layers.}
\label{Fig.4}
\end{figure}
\begin{figure}
\protect\caption{(a) Spatially resolved normalized conductance $(dI/dV)/(I/V)$ as a function of sample voltage in the vicinity of a ${\rm Te}_{\rm As}$ dopant atom. The conductance curves are calculated numerically after averaging about 70 measured $I(V)$ curves obtained within the three square areas which are indicated in the inset. The inset shows a tunneling current picture of 25 $\times$ 25 points of the dopant atom at a fixed tip-sample distance and at a sample voltage of $-1.5 \, {\rm V}$. The scanned area measures $5.8 \, {\rm nm} \times 5.8 \, {\rm nm}$. (b) Normalized conductance $(dI/dV)/(I/V)$ above a defect free area of the GaAs (110) surface for two different STM tips.}
\label{Fig.5}
\end{figure}
\begin{figure}
\protect\caption{(a) STM image of a ${\rm Te}_{\rm As}$ dopant atom taken at a sample voltage of $-1.5 \, {\rm V}$. The contribution of the atomic lattice has been filtered out. (b) Cross section of the STM topography along the white line indicated in (a).}
\label{Fig.6}
\end{figure}
\begin{figure}
\protect\caption{Schematic diagram of a nanoscale tunnel junction between a semiconductor surface and a metallic tip. $E_{C}$ and $E_{V}$ are the semiconductor conduction band edge and valence band edge, respectively. The width of the semiconductor band gap, $E_{C} - E_{V}$, is represented by the symbol $2\Delta$, while the conduction and valence band are assumed to have a width $4\Delta$. $\varepsilon_d$ corresponds to the (voltage dependent) position of an electron state which is localized within the tunnel junction area. $\Gamma$ is the tunneling rate, while $\gamma$ is the relaxation rate for the electrons in the semiconductor electrode and $\gamma_{0}$ is the relaxation rate for the localized state due to its hybridization with the metallic tip.}
\label{Fig.7}
\end{figure}
\begin{figure}
\protect\caption{Calculated tunneling conductivity curves obtained for different values of the parameters $\varepsilon_{d}^{0}$, $\Gamma$, $\gamma$ and $\gamma_{0}$ which characterize a nanoscale STM junction (see Fig.~\protect\ref{Fig.7}) . All quantities are measured in units of half the initial width of the semiconductor band gap. (1) $\varepsilon_{d}^{0} \, = \, -0.6\Delta$, $\Gamma \, = \, 0.01 \Delta$, $\gamma \, = \, 0.1 \Delta$, $\gamma_0 \, = \, 0.1\Delta$; (2) $\varepsilon_{d}^{0} \, = \, 2.4\Delta$, $\Gamma \, = \, 0.01 \Delta$, $\gamma \,=\, 0.25\Delta$, $\gamma_0 \, = \, 0.3\Delta$; (3) $\varepsilon_{d}^{0} \, = \, 2.6\Delta$, $\Gamma \, = \, 0.01 \Delta$, $\gamma \,=\, 0.05\Delta $, $\gamma_0 \, = \, 0.05\Delta $. The dotted line corresponds to the initial model density of states (DOS).}
\label{Fig.8}
\end{figure}
\end{document}